\documentclass[prl,aps,twocolumn,showpacs]{revtex4}

\usepackage{amssymb}
\usepackage{graphicx}
\def\figref#1{\mbox{Figure~\ref{#1}}}

\def\Figref#1{\mbox{Fig.~\ref{#1}}}
\def\Figrefs#1{\mbox{Figs.~\ref{#1}}}
\def\FIGref#1{\mbox{\ref{#1}}}
\def\Eqref#1{\mbox{Eq.~(\ref{#1})}}

\def\eqref#1{\mbox{(\ref{#1})}}

\begin{document}
\title{Logic-Memory Device of a Mechanical Resonator}

\author{Atsushi Yao} \email{yao@dove.kuee.kyoto-u.ac.jp} 
\affiliation{Department of Electrical Engineering, Kyoto University, Katsura, Nishikyo, Kyoto, 615-8510 Japan}
\author{Takashi Hikihara} \email{hikihara.takashi.2n@kyoto-u.ac.jp}
\affiliation{Department of Electrical Engineering, Kyoto University, Katsura, Nishikyo, Kyoto, 615-8510 Japan}

\date{\today}

\begin{abstract}
We report multifunctional operation based on the nonlinear dynamics in a single microelectromechanical system (MEMS) resonator. 
This Letter focuses on a logic-memory device that uses a closed loop control and a nonlinear MEMS resonator in which multiple states coexist. 
To obtain both logic and memory operations in a MEMS resonator, we examine the nonlinear dynamics with and without control input. 
Based on both experiments and numerical simulations, we develop a novel device that combines an OR gate and memory functions in a single MEMS resonator. 
\end{abstract}
\pacs{85.85.+j, 05.45.-a, 62.40.+i, 45.80.+r,}

\maketitle 
Microelectromechanical systems or nanoelectromechanical systems (MEMS or NEMS) resonators 
have been developed for use as filters, frequency references, and sensor elements~\cite{V.Kaajakari2009}. 
Recently, significant research has focused on mechanical computation based on MEMS or NEMS resonators~\cite{Badzey2004, badzey2005coherent, Masmanidis2007, Mahboob2008, guerra2010noise, Noh2010, Unterreithmeier2010b, Mahboob2011, hatanaka2012electromechanical, Yao2012, Atsushi2012, khovanova2012minimal, 
wenzler2013nanomechanical, uranga2013exploitation, yao2013counter, mahboob2014multimode}. 
Some studies have shown that a single mechanical resonator can be used as a mechanical 
1-bit memory~\cite{Badzey2004,  badzey2005coherent, Mahboob2008, Noh2010, Unterreithmeier2010b, Yao2012, Atsushi2012, khovanova2012minimal, uranga2013exploitation} 
or as mechanical logic gates~\cite{guerra2010noise, Mahboob2011, hatanaka2012electromechanical}. 
Recently, multifunctional operation has been demonstrated in the form of a shift-register and a controlled NOT gate 
made from a single mechanical resonator~\cite{mahboob2014multimode}. 
The next phase is to use a closed loop control to generate multifunction devices, which consist of memory and multiple-input gates, in a single device.
The closed loop allows output and excitation signals to be fixed at a single frequency. 
The goal of the work presented in this Letter is to develop multifunction operation from a nonlinear 
MEMS resonator in which multiple states coexist with closed loop control. 

Nonlinear dynamical responses are commonly observed in a MEMS resonator. 
The nonlinear dynamics of the MEMS resonator is well known to be described by the Duffing 
equation~\cite{Badzey2004, Mestrom2008, Unterreithmeier2010b, Naik2012, antonio2012frequency, Atsushi2012}. 
Such a nonlinear MEMS resonator has hysteretic characteristics, which lead to two stable states and one unstable state, 
depending on the frequency~\cite{V.Kaajakari2009, JohnA.Pelesko2003} 
or excitation force~\cite{guerra2010noise, uranga2013exploitation}. 

This Letter focuses on fabricating a multifunction device that offers logic and memory (called a ``logic-memory device''). 
To do this, we examine the nonlinear dynamics in a MEMS resonator with and without control input. 
In the following, we discuss the experiments and numerical simulations 
that allowed us to develop a device that combines multiple-input gate and memory functions in a single nonlinear MEMS resonator.

\begin{figure}[!b]
 \begin{center}
 \includegraphics[width=1\linewidth]{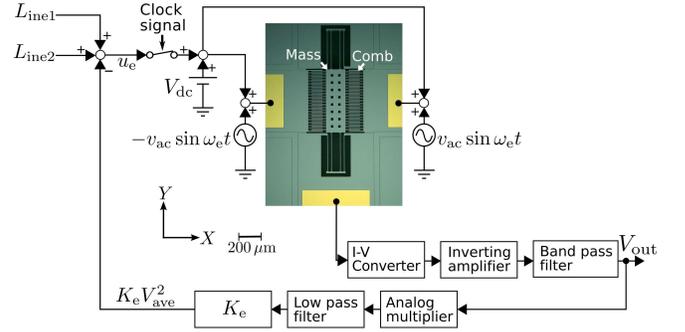}
\caption{Schematic of MEMS resonator, measurement system, and control system that relates to logic inputs. 
The nonlinear MEMS resonator, fabricated using silicon-on-insulator technology, 
is actuated by an ac excitation voltage $v_{\rm ac}$ with a dc bias voltage $V_{\rm dc}$. 
When the MEMS resonator is excited, the mass vibrates in the $X$-direction. 
In the measurement system, 
the output voltage $V_{\rm out}$ depends on the amplitude and phase of the displacement in the nonlinear MEMS 
resonator. 
The control system is implemented with a feedback input and logic inputs. 
The feedback input is given as the slowly changing dc voltage $V_{\rm ave}^{2}$, 
to which the output voltage $V_{\rm out}$ is converted by an analog multiplier and a low-pass filter (see Ref.~\cite{Yao2012} for more details.). 
The logic inputs, represented by two dc voltages ($L_{\rm ine 1}$ and $L_{\rm ine 2}$), 
are added to the dc bias voltage $V_{\rm dc}$. 
As a result, the excitation force under control becomes proportional to 
$V_{\rm dc} + u_{\rm e} = V_{\rm dc} + L_{\rm ine 1} + L_{\rm ine 2} - K_{\rm e} V_{\rm ave}^{2}$. 
Here $L_{\rm ine 1}$ and $L_{\rm ine 2}$ denote the input signals, which serve as the logic inputs, $u_{\rm e}$ is the control input, and 
$K_{\rm e}~( = 11)$ is the feedback gain in the experiments. 
}
\label{fig1}
 \end{center}
\end{figure}

The proposed comb-drive MEMS resonator is shown in \Figref{fig1}. 
The resonator consists of a perforated mass with a width, length, and thickness of $175$, $575$, and $25$\,$\mu$m, 
respectively~\cite{Naik2012, Naik2011, Naik2011a}. 
When the comb-drive resonator is electrically excited, the mass vibrates in the lateral direction.
The vibration of the mass is detected by using 
a differential measurement~\cite{V.Kempe2011} in vacuum (around $10$\,Pa) at room temperature. 
The output voltage of the differential measurement is $V_{\rm out} \propto  v_{{\rm ac}} A_{\rm e} \sin (2 \omega_{\rm e} t + \phi_{\rm e})$, 
where $A_{\rm e}$ denotes the displacement amplitude and $\phi_{\rm e}$ is the phase. 
The vibration displacement is measured without additional sensors; therefore, the MEMS resonator is equipped with a 
comb drive that normally serves as a forcing actuator, but which simultaneously serves as a displacement sensor~\cite{Yao2012,yao2013counter}. 

\begin{figure}[!t]
 \begin{center}
\includegraphics[width=1\linewidth]{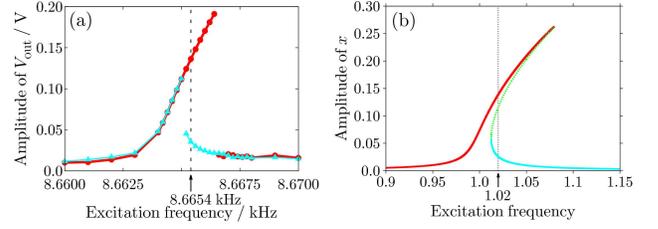}
 \caption{(color online) (a) Experimentally obtained frequency response curves at $V_{\rm dc} = 150$\,mV and $u_{\rm e} = 0.0$\,mV. 
The dark (red) and thin (aqua) lines correspond to the responses to increasing and decreasing frequency sweeps, respectively. 
In the hysteresis region, two coexisting stable states appear that strongly depend on the sweep direction. 
(b) Corresponding numerical response curves generated from \Eqref{eq:hou} for $k_{\rm n} = 0.001$ and $u_{\rm n} = 0.0$. 
The solid (red and aqua) lines show two stable solutions and the dashed (green) line shows an unstable solution. 
} 
\label{fig2}
  \end{center}
\end{figure}

\figref{fig2}(a) shows the amplitude frequency response (without control input $u_{\rm e}$). 
The MEMS resonator produces a hysteretic response: the curves differ for increasing and decreasing frequency sweeps. 
The nonlinear dynamics of the MEMS resonator is qualitatively modeled by the non-dimensional equation as follows: 
\begin{eqnarray}
\frac{{\rm d^{2}}x }{{\rm d}t^{2}} +  \frac{1}{Q}\frac{{\rm d}x}{{\rm d}t}  + x + \alpha_{3} x^3 &=& (k_{\rm n} + u_{\rm n}) \sin \omega_{\rm n} t {\rm ,}  \label{eq:hou}
\end{eqnarray}
where $x$ denotes the displacement, $\omega_{\rm n}$ is the excitation frequency, 
$Q~(= 282)$ is the quality factor, $\alpha_{3}~(= 3.23)$ is 
the nonlinear mechanical spring constant, $k_{\rm n}$ is the amplitude of the excitation force, 
which in our experiments is proportional to a dc bias voltage $V_{\rm dc}$, and $u_{\rm n}$ is the control input. 
The parameter settings are obtained from Ref.~\cite{Naik2012}, which deals with the same device as depicted in \Figref{fig1}. 
\Figref{fig2}(b) shows the amplitude as a function of excitation frequency for the resonator as determined by numerical simulations 
at $k_{\rm n} = 0.001$ and $u_{\rm n} = 0.0$. 
At any given frequency in the hysteretic region, the MEMS resonator exhibits two coexisting stable states. 
In the following experiments (simulations), the excitation frequency is fixed at $8.6654$\,kHz ($1.02$). 

\begin{figure}[!t]
 \begin{center}
\includegraphics[width=1\linewidth]{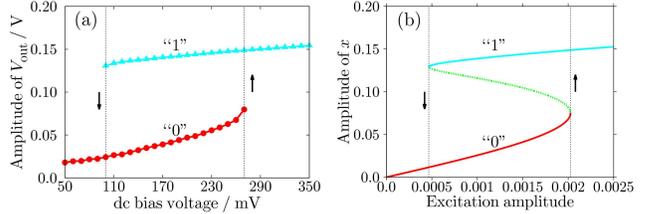}
 \caption{(color online) (a) Measured hysteretic characteristics with respect to dc bias voltage $V_{\rm dc}$ at $8.6654$\,kHz and $u_{\rm e} = 0.0$\,mV. 
The excitation amplitude in the absence of the control input is proportional to the dc bias voltage $V_{\rm dc}$. 
The experimentally obtained response shows the hysteretic behavior 
when the excitation amplitude is swept from left to right (thick red line) and right to left (thin aqua line). 
The hysteresis region exists at $90$\,mV $< V_{\rm dc} < 280$\,mV. 
For logic and memory output, the thin (dark) line is regarded as a logical ``$1$'' (logical ``0''). 
(b) Corresponding numerical hysteretic characteristics as a function of 
excitation amplitude $k_{\rm n}$ at $\omega_{\rm n}=1.02$. 
The solid (red and aqua) lines show stable regions and the dashed (green) line shows an unstable region. 
These stable regions can be used as two states, corresponding to logical ``0'' and ``1'', 
for logic and memory functions, as in panel (a). 
} 
\label{fig3}
  \end{center}
\end{figure}

\figref{fig3}(a) shows the experimentally determined hysteretic behavior as a function of dc bias voltage $V_{\rm dc}$ at $u_{\rm e} = 0.0$\,mV. 
The corresponding numerical results are shown in \Figref{fig3}(b) as a function of excitation amplitude $k_{\rm n}$. 
The numerical results are consistent with the experimental results. 
The nonlinear MEMS resonator has stable regions (solid line) that are completely separated by an unstable region (dashed line). 
These stable regions, which correspond to large and small amplitude vibrations, 
define the two states of the single-output logic or memory device in a single MEMS resonator. 
In the numerical simulations (experiments), a displacement amplitude greater than 0.1 (output voltage amplitude greater than 0.11 V) 
is regarded as a logical ``1''; a value less than 0.1 (0.11 V) is regarded as a logical ``0'' for logic and memory output. 
Hereinafter, the excitation force $k_{\rm n}$ is fixed at $0.001$ in the numerical simulations. 
In the corresponding experiments, the dc bias voltage $V_{{\rm dc}}$ and the ac excitation amplitude $v_{{\rm ac}}$ are set to $150.0$\,mV 
and $0.6$\,V, respectively. 

We now discuss the nonlinear dynamics with control input as a logic operation. 
\Figref{fig1} shows the control system to perform the logic operation. 
The switching between two coexisting stable states was done 
by a displacement feedback control in the nonlinear MEMS resonator~\cite{Yao2012}. 
Based on the results, the feedback control is performed. 
The logic inputs 
are applied to the MEMS resonator in the form of two dc voltages ($L_{\rm ine 1}$ and $L_{\rm ine 2}$). 
The control input $u_{\rm e}$ is described as follows: 
\begin{eqnarray}
u_{\rm e} &=& L_{\rm ine 1} + L_{\rm ine 2} - K_{\rm e} V_{\rm ave}^{2}, 
\end{eqnarray}
where $K_{\rm e}$ denotes the feedback gain and $V_{\rm ave}^{2}$ a slowly changing dc voltage that depends 
on the displacement~\cite{Yao2012}. 

For this experimental method, the control input $u_{\rm n}$ is 
described as follows~\cite{Atsushi2012}: 
\begin{eqnarray}
u_{\rm n} &=& L_{\rm in n 1} + L _{\rm in n 2} - K_{\rm n}  A^{2}_{\rm {\rm n}ave}  {\rm ,} \\
 A^{2}_{\rm {\rm n}ave} &=&  \frac{ A^{2}_{{\rm n}1} +  A^{2}_{{\rm n}2} + 	\cdot 	\cdot 	\cdot + A^{2}_{{\rm n}m} +	\cdot 	\cdot 	\cdot  + A^{2}_{{\rm n}M} }{M} {\rm ,} 
\end{eqnarray}
where $L_{\rm in n 1}$ and $L _{\rm in n 2}$ denote the input signals that are the logic inputs, 
$K_{\rm n}$ is the feedback gain, $m$ is a natural number, 
$M$ is the average number, and $A_{{\rm n}m}$ is the displacement amplitude of the previous $m$ period within $1 \leq m  \leq M$ 
for the numerical simulations. 
In this case, $A^{2}_{\rm {\rm n}ave}$ is the average of $A^{2}_{{\rm n}m}$. $K_{\rm n}$ is set to $0.08$ and $M$ is set to $100$. 

\begin{figure}[!t]
 \begin{center}
\includegraphics[width=1\linewidth]{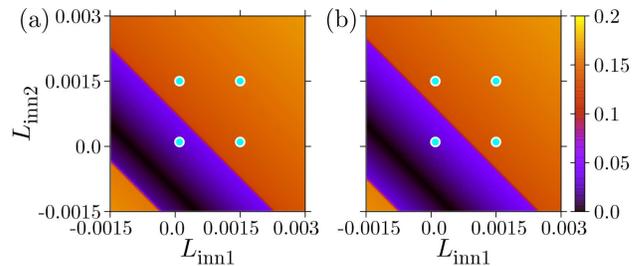}
 \caption{(color online) 
Amplitude modulation systematically varied in input signals $L_{\rm inn1}$ and $L_{\rm inn2}$ at $\omega_{\rm n} = 1.02 $ and $k_{\rm n}= 0.001$ 
(numerical results). 
The light (dark) region corresponds to more than (less than) 0.1 in displacement amplitude, corresponding to a logical ``1'' (logical ``0'') output. 
For an OR gate, input signals are set to $0.0015$ and $0.0001$, as shown by the four circles with light (aqua) color: 
(a) Initial state is set to the small amplitude solution (logical ``0'' for memory output) 
(b) Initial state is the large amplitude solution (logical ``1'' for memory output).
} 
 \label{fig4} 
\end{center}
\end{figure}

\figref{fig4} shows the numerically obtained steady states when the control input is applied to the MEMS resonator. 
The control input can induce a modulation of the resonator's amplitude 
and thus change the logical value of the output. In \Figref{fig4}(a) (\Figref{fig4}(b)), the initial state of memory output is a logical ``1'' 
(logical ``0''). 
In \Figrefs{fig4}(a) and \FIGref{fig4}(b), 
there exist regions in which the displacement amplitude is the same because the control system depends on the feedback input. 
Assume that the control system receives just the input signals $L_{\rm in n 1}$ and $L _{\rm in n 2}$. 
When the input signals $L_{\rm in n 1} = L _{\rm in n 2}= 0.0$ are sent to the MEMS resonator, 
the large stable state cannot switch to the small state. 
Therefore, a single MEMS resonator can be used as a logic gate because of the 
adjustment of the logic inputs and the feedback input. 

\begin{figure}[!t]  
 \begin{center}
\includegraphics[width=1\linewidth]{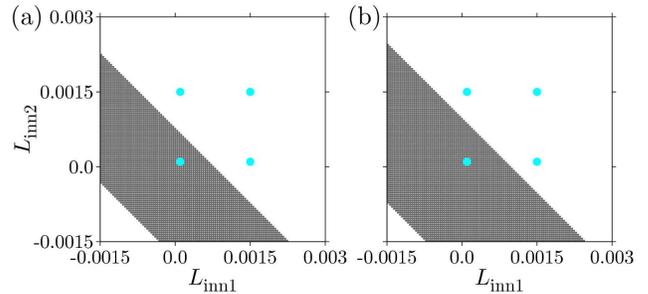}
 \caption{(color online) 
Calculated convergence conditions (final state) when the control input shown in \Figref{fig4} is off. 
The white (black) region corresponds to convergence to a logical ``1'' (logical ``0'') for memory output. 
Light (aqua) circles show the logic input in our simulations, as in \Figref{fig4}: 
(a) Numerical results corresponding to \Figref{fig4}(a). 
(b) Numerical results corresponding to \Figref{fig4}(b). 
} 
 \label{fig5}
\end{center}
\end{figure}

To execute a memory operation in a MEMS resonator, 
we must consider the nonlinear dynamics without the control input. 
When the control input is not applied, 
every initial state corresponds to the convergence to either the small amplitude (black) 
or large amplitude (white) solutions, as shown in \Figref{fig5}. 
In the nonlinear MEMS resonator, the small and large amplitude solutions have each basin of attraction~\cite{Guckenheimer1983} 
around each solution. 
The convergence conditions depend on the two basins of attraction~\cite{Unterreithmeier2010b, Atsushi2012}. 
The light region (displacement amplitude greater than 0.1) in \Figref{fig4} corresponds to white region in \Figref{fig5} and vice versa. 
When the control input is not applied, the MEMS resonator maintains its original logic output. 
Thus, the nonlinear MEMS resonator works as the memory device by storing the logic information. 

The results shown in \Figrefs{fig4} and \FIGref{fig5} indicate that MEMS resonator works as a combined logic-memory device. 
The logic and memory operations 
can be programmed by adjusting the resonator's operating parameters (input signals). 
In the numerical simulations, when the input signal ($L_{\rm inn 1}$ or $L_{\rm inn 2}$) 
is set to $0.0015$ ($0.0001$), the logic input is regarded as logical $1$ (logical $0$). 
The logic inputs (0, 0) of input signals ($L_{\rm inn 1}$, $L_{\rm inn 2}$) 
have a value of $0.0002$, (0, 1) and (1, 0) have a value of $0.0016$, and finally (1, 1) have $0.0030$, as shown by the light (aqua) circles in 
\Figrefs{fig4} and \FIGref{fig5}. 
The output of the device is a logical ``0'', 
when the logic inputs are (0, 0), which correspond to a value of $0.0002$. 
However, when the logic inputs are set to (0, 1), (1, 0), or (1, 1), the output corresponds to a logical ``1''. 
Therefore, the single MEMS resonator combines the function of an OR gate and memory. 

\begin{figure}[!t] 
\begin{center}
\includegraphics[width=1\linewidth]{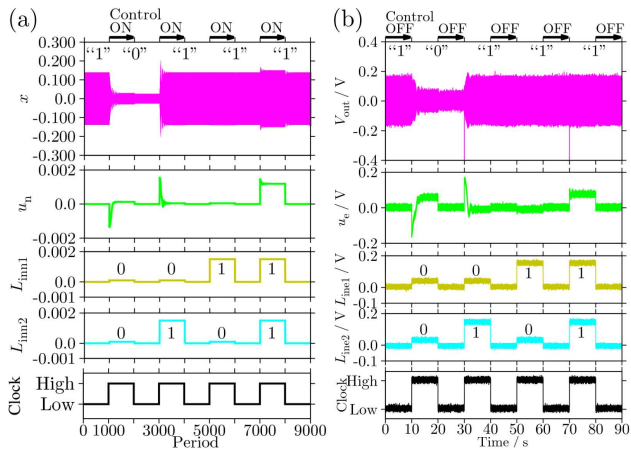}
 \caption{(color online) 
Time evolution of the combined device (OR gate and memory). The initial memory output is a logical ``1''. 
When the clock signal is high (low), the control input is (is not) applied to the MEMS resonator. 
Note that the nonlinear MEMS resonator is used as a logic (memory) device at high (low) clock signal. 
The logic inputs start from (0, 0) and continue to (0, 1), (1, 0), and (1, 1). 
The logic output in a MEMS resonator changes to logical ``0'', ``1'', ``1'', and ``1'' at each high clock signal: 
(a) Numerical results. 
(b) Experimental results. 
In our experiments, the logic inputs (0, 0) 
of experimental input signals ($L_{\rm ine1}$, $L_{\rm ine2}$) have a voltage of $75.0$\,mV, (0, 1) and (1, 0) 
have a voltage of $187.5$\,mV, and finally (1, 1) have $300.0$\,mV. 
}
\label{fig6}
\end{center}
\end{figure}

These logic and memory operations can be demonstrated experimentally in a single MEMS resonator. 
The operations are confirmed for the behavior of device at clock evolution. 
The calculated time evolutions of the device are shown in \Figref{fig6}(a) and the corresponding experimental time evolutions are shown 
in \Figref{fig6}(b). 
The calculated results are qualitatively consistent with the experimental results. 
When electrical noise and/or surges appear in the experiments, 
no logic faults occur and the memory operations are not perturbed. 
In the experiments, the feedback gain and input signals are swept within the operating range and adjusted. 
The experimental modulation of the amplitude and the convergence conditions will be examined in more detail in a future presentation. 
Nevertheless, this work demonstrates both 
experimentally and numerically a combined OR gate and memory functions in a single MEMS resonator.

In conclusion, 
we numerically and experimentally demonstrated a multifunctional device consisting of a nonlinear MEMS resonator. 
We confirmed that when a control input is applied to a nonlinear MEMS resonator, two equal-amplitude regions exist because of the 
adjustment of the feedback input. 
Therefore, a single MEMS resonator can work as an OR gate. 
We also used numerical simulations to show that 
in the absence of the control input, the nonlinear MEMS resonator maintains its original logical state. 
Thus, this resonator also serves as a memory device. 
Therefore, we demonstrate a novel combination of an OR gate and a memory device in a single MEMS resonator. 
By considering the closed loop, these results open the way to further research in multifunctionality in the nonlinear MEMS resonator, 
which may take the form of multiple-input gates such as three- or four-input logic gates and memory.

We are grateful to Dr.~S.~Naik (SPAWAR, USA) for his support in the design of MEMS resonators. 
This work was partly supported by the Global COE of Kyoto University, Regional Innovation Cluster Program 
``Kyoto Environmental Nanotechnology Cluster'', and the JSPS KAKENHI (Grant-in-Aid for Exploratory Research) $\sharp$21656074. 
A.Y. acknowledges support from the Japan Society for the Promotion of Science.

\end{document}